\documentclass[amsmath, amssymb,aps, pre,twocolumn]{revtex4-1}
\usepackage[latin1]{inputenc}
\usepackage{graphicx}
\usepackage{color}
\usepackage{theorem}

\renewcommand{\v}[1]{{\boldsymbol #1}}

\DeclareGraphicsExtensions{.pdf}
\graphicspath{{./figs/},{../sem/figs/}}

\theorembodyfont{\normalfont}
\newtheorem{theorem}{Theorem}

\begin{document}
\pagestyle{plain}
\title{Equivalence of Deterministic walks on regular lattices on the plane}
\author{Ana Rechtman}
\affiliation{IRMA, Universit\'e de Strasbourg,\\ 7 rue Ren\'e
  Descartes,  67084 Strasbourg, France}
\email{rechtman@math.unistra.fr}

\author{Ra\'ul Rechtman}
\affiliation{Instituto de Energ\'i{}as Renovables, 
 Universidad Nacional Aut\'onoma de M\'exico,\\  
 Apdo.\ Postal 34, 62580 Temixco Mor., M\'exico}
\email{rrs@ier.unam.mx}
 
\begin{abstract}
  We consider deterministic walks on square, triangular and hexagonal
  two dimensional lattices. In each case, there is a scatterer at
  every site that can be in one of two states that force the walker to
  turn either to his/her immediate right or left. After the walker is
  scattered, the scatterer changes state. A lattice with an
  arrangement of scatterers is an environment. We show that there are
  only two environments for which the scattering rules are injective, mirrors or
  rotators, on the three lattices. On hexagonal lattices, B. Z. Webb
  and E. G. D. Cohen~\cite{webb2014},
  proved that given an initial position and velocity of the
  walker and an environment of one type of scatterers, mirrrors or
  rotators, there is an environment of the other type such that the
  walks on both environments are equivalent, meaning they visit the
  same sites at the same time steps.  We prove the equivalence of
  walks on square and triangular lattices and include a proof of the
  equivalence of walks on hexagonal lattices.  The proofs are based
  both on the geometry of the lattice and the structure of the
  scattering rule.
\end{abstract}


\maketitle

\section{Introduction}
\label{intro}

In 1912 Paul and Tatiana Ehrenfest published a monograph on the
foundations of statistical mechanics in the {\em Encyklop\"adie der
  Mathematische Wissenschaften} where among other subjects they
presented the wind-tree model to explain Boltzmann's transport
equation~\cite{ehrenfest1912}. The wind is formed by particles that do
not interact between themselves and move with the same speed along the
$x$ and $y$ axes. The trees are fixed squares randomly placed
  on the plane with their diagonals aligned along the $x$ and $y$ axes
  that scatter the wind particles an angle of $\pm\pi/2$.  The motion
of the wind particles is deterministic and time reversible.  The model
is a Lorentz gas so one can consider only one particle. The wind-tree
model served as a starting point for the study of Lorentz lattice
gases where a particle advances from a site to one of its nearest
neighbor sites on a lattice in one time step and the trees, or better
scatterers, occupy one site. Following Bunimovich we speak of walks on
an environment of scatterers~\cite{bunimovich04}.  On a square lattice
$\mathbb{Z}^2$ the trees became two types of mirrors, small line
segments at angles of $\pi/4$, a right mirror, and $3\pi/4$, a left
mirror, with respect to the positive $x$ axis.  In the flipping model,
the scatterers flip from one orientation to the other one, after the
particle is scattered~\cite{ruijgrok88,cohen92}. Right (left) rotators
have also been studied where the walker is scattered an angle of
$\pi/2$ to his/her right (left) on a square
lattice~\cite{gunn85,langton86,meng94}.  Walks on two dimensional
triangular lattices
$\mathbb{T}^2$~\cite{kong91,cohen95,cohen95a,bunimovich92,bunimovich04}
and on hexagonal lattices
$\mathbb{H}^2$~\cite{wang95,webb2014,webb2015} have been studied
extensively.

We consider deterministic discrete walks on two dimensional regular
lattices where the walker moves with unit speed from a site to a
nearest neighbor site in one time step.  All the sites are occupied by
a scatterer that flips
from right to left and vice versa after the
walker is scattered. A walker on a flipping mirror environment in
$\mathbb{Z}^2$ at a site $(x,y)$ is reflected by a mirror an angle of
$\pm\pi/2$ and jumps with unit speed to one of two  nearest
neighbor sites of $(x,y)$. The mirror at $(x,y)$ changes orientation
as the walker passes by rotating an angle of $\pi$.  In a rotator
environment in $\mathbb{Z}^2$, the walker at $(x,y)$ turns an angle of
$\pi/2$ to his/her right (left) when there is a right (left) rotator
at $(x,y)$ and jumps to the nearest neighbor site in front of him/her.
The rotator at $(x,y)$ changes from right to left or left to right as
the walker passes.  The change in orientation of the mirrors or
rotators, referred to as a flip, forbids the existence of closed
orbits~\cite{bunimovich92}.

\begin{figure}
  \begin{center}
    (a)\\
    \hskip -2mm\includegraphics[width=0.77\columnwidth]%
                               {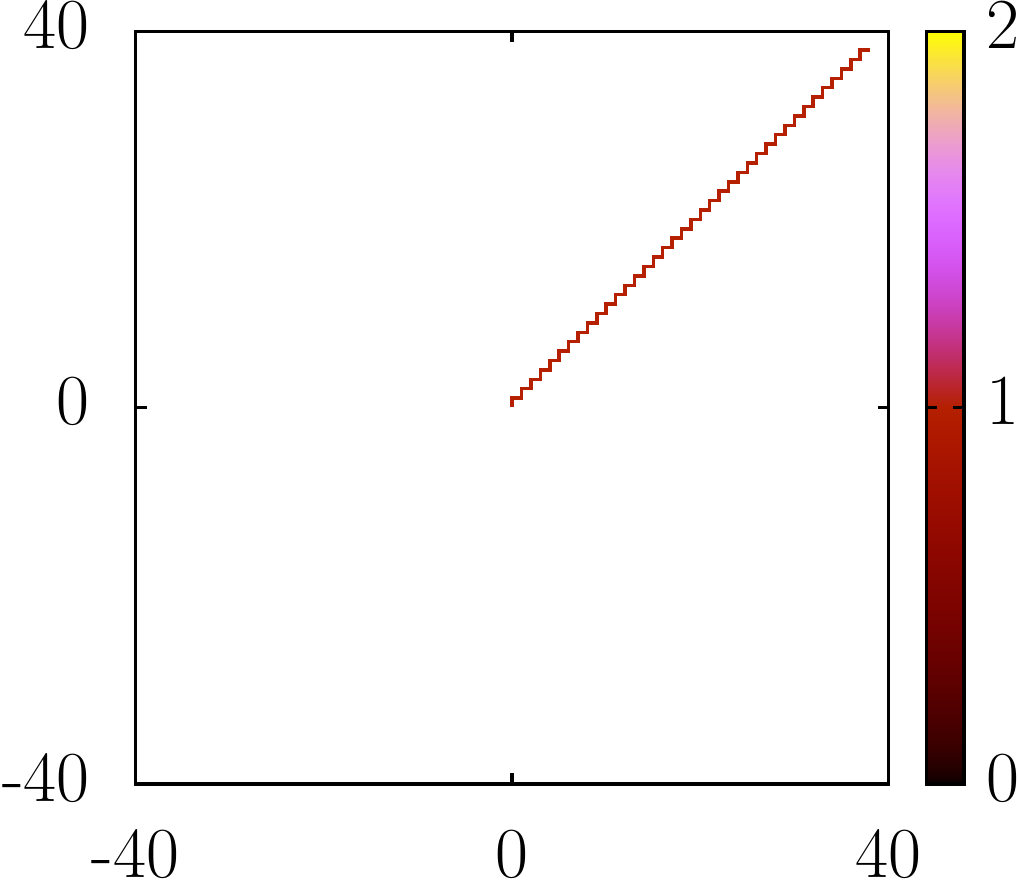}\\[3mm]
    (b)\\
    \includegraphics[width=0.8\columnwidth]{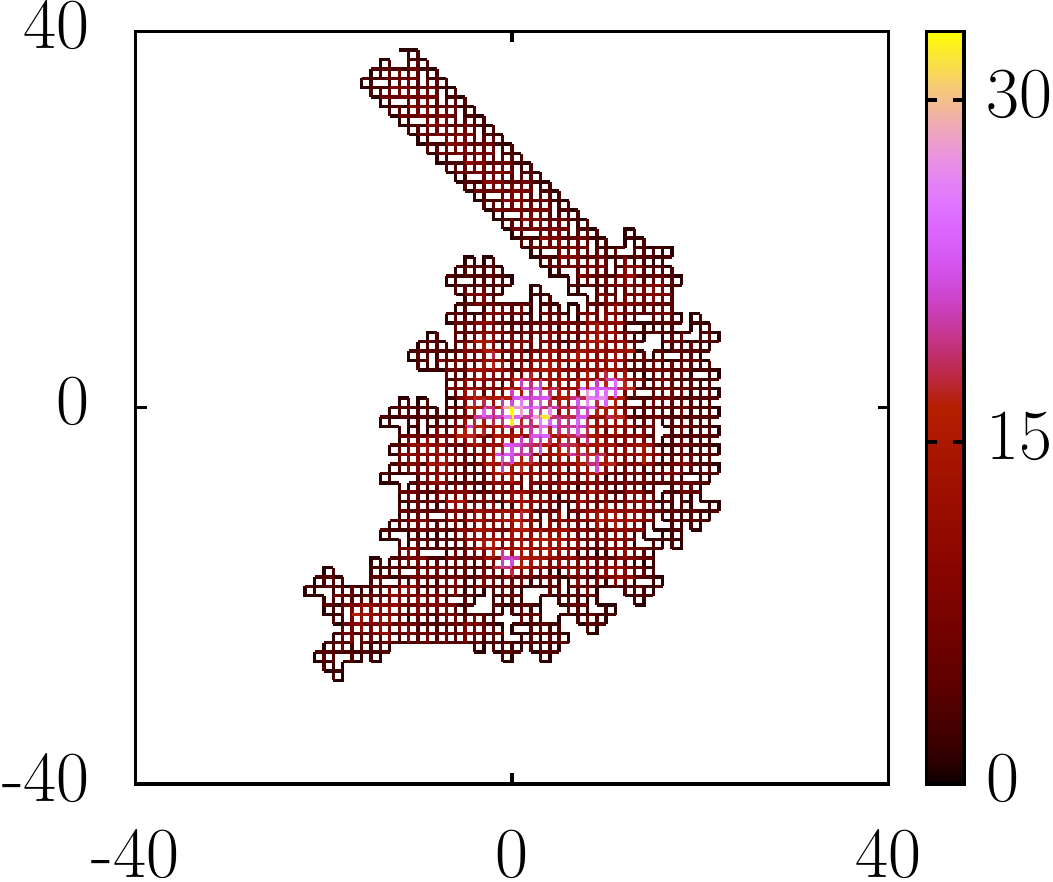}  
    \caption{\label{fig:FM-FR} (Color online).  (a) A walk on an
      initially ordered flipping mirror environment of right mirrors
      in $\mathbb{Z}^2$. The walker alternates between a vertical and
      a horizontal step. (b) A walk on an initially ordered flipping
      rotator environment of right rotators in $\mathbb{Z}^2$, showing
      the visited sites. After 9,977 time steps the walker moves on a
      ``highway'' advancing two sites horizontally and two vertically
      every 104 time steps. In (a) and (b), at time $t=0$ the walker is at
      the origin with velocity $(1,0)$. (The color scale indicates the
      number of times the walker is at a site.) }
 \end{center}    
\end{figure}

A walker on an environment
initially filled with right mirrors in $\mathbb{Z}^2$ moves in a
zigzag, alternating between a vertical step and a horizontal one
with a speed of $\sqrt{2}/2$, as we show in Fig.~\ref{fig:FM-FR} (a). The
initial velocity of the walker determines on which of the four
diagonals he/she will move.  On the other hand, a walker on an
environment initially filled with right rotators in $\mathbb{Z}^2$
moves around his/her starting point and after 9,977 time steps,
advances two sites horizontally and two vertically every 104 time
steps in what is known as a highway with a speed of $\sqrt{2}/52$ as
we show in Fig.~\ref{fig:FM-FR}
(b)~\cite{langton86,gale95,boon01,meng94}. The initial velocity of the
walker determines the direction of the highway.

\begin{figure}
  \begin{center}
    (a)\\
    \includegraphics[width=0.8\columnwidth]%
                                {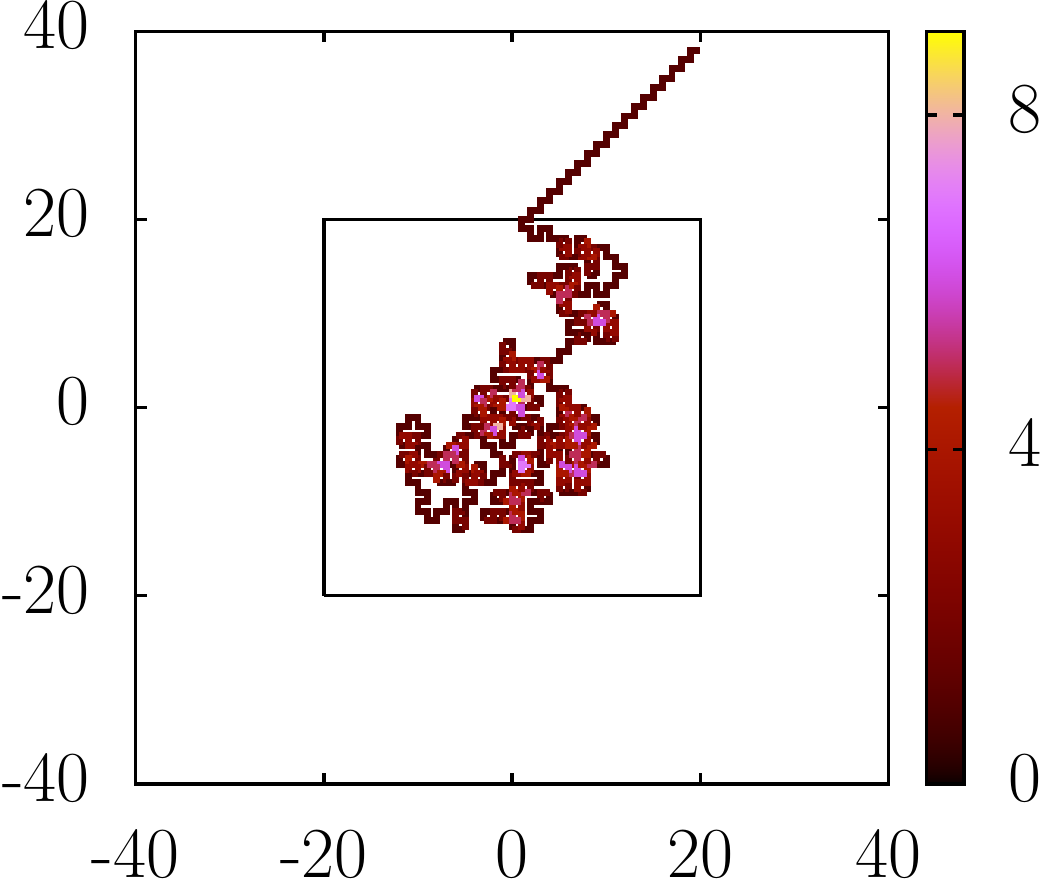}\\[3mm]
    (b)\\
    \includegraphics[width=0.8\columnwidth]{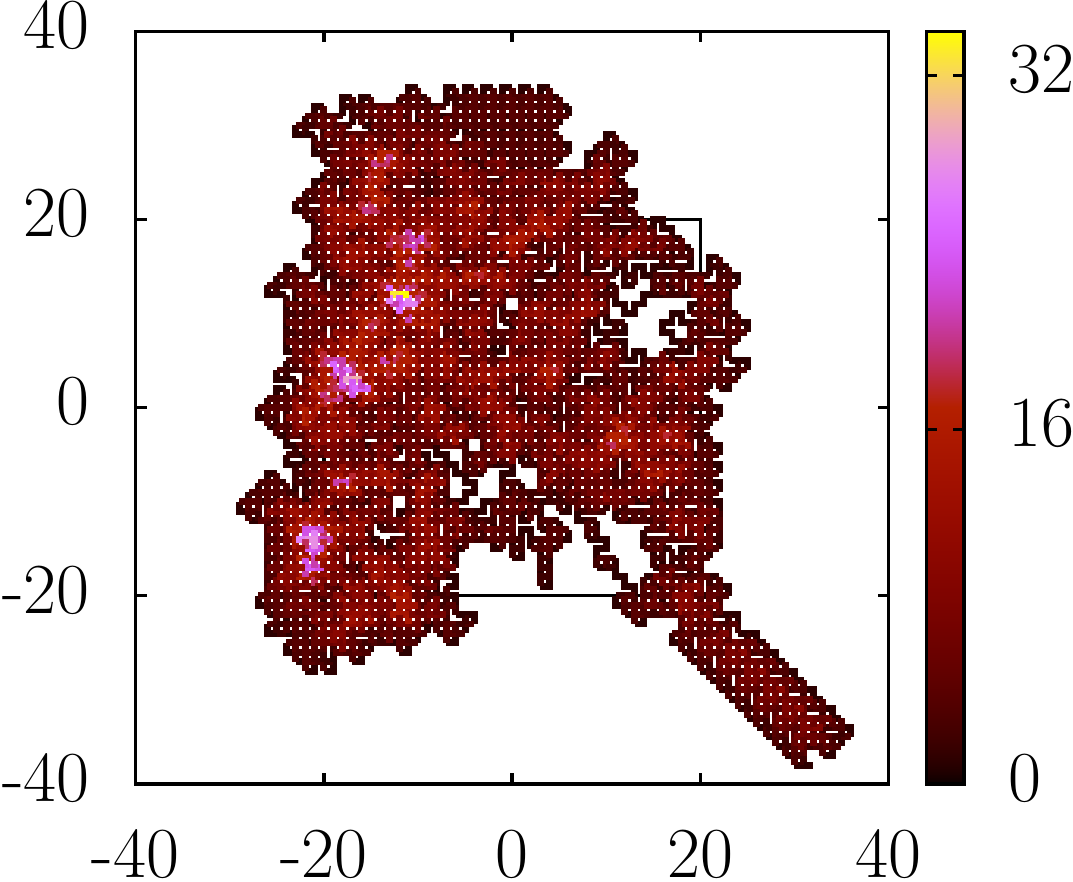}  
    \caption{\label{fig:disorder} (Color online).  In (a) and (b), at
      time $t=0$, the environment is initially ordered with right
      scatterers outside the square of side 40 and inside this square
      the probability that any given site has a right scatterer is
      1/2. Initially, the walker is at the origin with velocity
      $(1,0)$. (a) A walk on a flipping mirror environment. (b) A walk on a flipping
      rotator environment.  (The color scale indicates the number of
      times the walker is at a site.) }
 \end{center}    
\end{figure}

Walks on both environments are also different on initially disordered
environments. What is maybe more striking is that if the environment
is disordered in some region and ordered in another one as in
Fig.~\ref{fig:disorder}, the walker in the mirror environment will
eventually leave the disordered region and move alternatively one site
horizontally and the next vertically, Fig.~\ref{fig:disorder} (a).
The walker in the rotator environment will eventually leave the
disordered region and walk on a highway, Fig.~\ref{fig:disorder} (b).

The previous Figs. show that in $\mathbb{Z}^2$, walks on mirror
environments are very different from from those on rotator
environments. For walks in triangular lattices, $\mathbb{T}^2$, mirror
and rotator scatterers can be defined but walks on the two
environments are not so strikingly different. Again, mirror and
rotator scatterers can be defined for walks in hexagonal lattices,
$\mathbb{H}^2$, and
walks on both environments are very different. 

In this paper we are interested in the scattering rules of flipping
mirror and flipping rotator environments in the three types of
lattices, as defined in the literature.  As we prove, the choice is
justified by the fact that these are the only injective scattering
rules meaning that walkers arriving to a scatterer with different
velocities, will leave the site with different velocities.  In the
case of a flipping mirror environment on $\mathbb{Z}^2$ and
$\mathbb{T}^2$ the walks are time reversible if in the time reversed
walk the scatterer flips before scattering the walker, and in the
other cases, walks are time reversible if in the time reversed walk
the scatterer flips after the walker is scattered.

Given an initial position and velocity of a walker on one environment
with one type of scatterers in $\mathbb{H}^2$, B. Z. Webb and
E. G. D. Cohen proved that there is an environment with the other type
of scatterers such that the walks on both environments are equivalent
in the sense that their trajectories,
$\{(x(t),y(t))\,|\,t\in\mathbb{N}\}$ are the same~\cite{webb2014}.
Also, L. A. Bunimovich and S. E. Troubetzkoy stated the equivalence of
walks on mirror and rotator environments with scatterers that do not
flip in $\mathbb{Z}^2$, \cite{bunimovich92}. Our aim is to prove the
equivalence of walks on square $\mathbb{Z}^2$ and triangular
$\mathbb{T}^2$ lattices with flipping scatterers.

In Sec.~\ref{sec:square} we present the walks on mirror and rotator
environments in $\mathbb{Z}^2$, show that these are the only
injective scattering rules, and prove the equivalence of walks as stated above.
In the next Sec. we prove the same results for walks in $\mathbb{T}^2$ and for completeness, in
Sec.~\ref{sec:hexagon}, we also prove the equivalence of walks in
$\mathbb{H}^2$. The proofs are based on an interplay between the geometry of
the environment and the scattering rule.  We close with some
conclusions.


\section{Walks in ${\mathbb Z}^2$}
\label{sec:square}

The main result of this Sec. is that given the initial position and
velocity of a walker on an environment of scatterers, mirrors or
rotators, in $\mathbb{Z}^2$, there is an environment with the other type of
scatterers such that both walks are equivalent. We start with some
definitions, then illustrate and prove the result.

\begin{figure}
 \begin{tabular}{cc}
  (a) & \hskip 5mm (b)\\[2mm]
  \includegraphics[width=0.4\columnwidth]{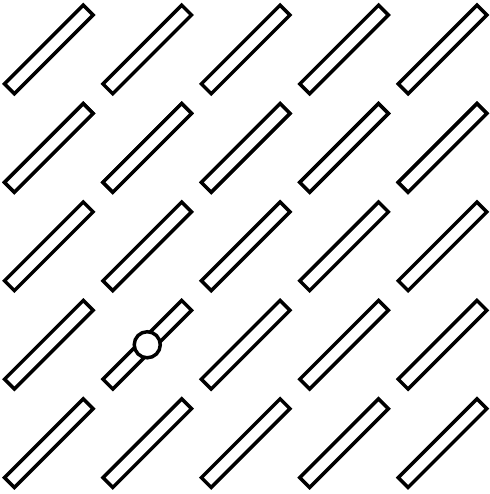} &
  \hskip 5mm\includegraphics[width=0.4\columnwidth]{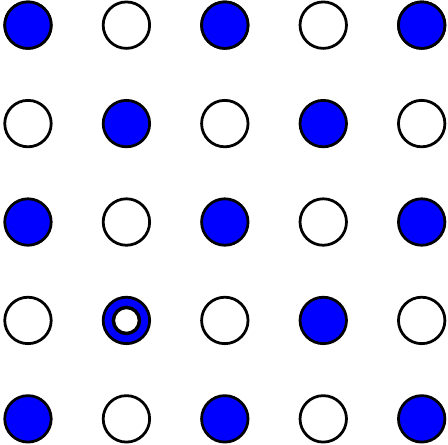} \\[1mm]
  (c) & \hskip 5mm (d)\\[2mm]
  \includegraphics[width=0.4\columnwidth]{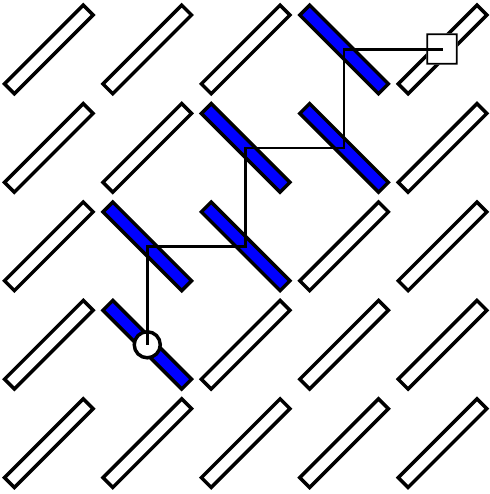} &
  \hskip 5mm\includegraphics[width=0.4\columnwidth]{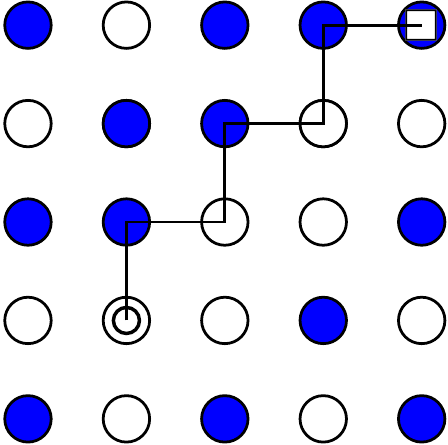} 
 \end{tabular}
 \caption{\label{fig:FMOE-FRCE} (Color online). (a) Initially ordered
   flipping mirror environment, $OFME$. (b) Initially 
   checkerboard flipping rotator environment, $CFRE$. (c) A
   walk on the $OFME$ at $t=6$ showing the walk and the
   state of the environment.  (d) A walk on the $CFRE$  at
   $t=6$ showing the walk and the state of the environment. In both
   cases, at $t=0$ the walker is at the site marked by a small open
   circle and has velocity $\v{v}_0=(1,0)$ and at $t=6$, the walker's
   position is marked by a small open square. Right mirrors make an
   angle of $\pi/4$ with the positive $x$ axis, shown as open
   rectangles (in white).  Left mirrors make an angle of $3\pi/4$
   with the positive $x$ axis, shown as dark rectangles (in
   blue). Right and left rotators are shown as open circles (in white)
   and dark circles (in blue) respectively.}
\end{figure}

\begin{figure}
 \begin{center}
 \begin{tabular}{cc}
  (a) & \hskip 5mm (b)\\[2mm]
   \includegraphics[width=0.4\columnwidth]{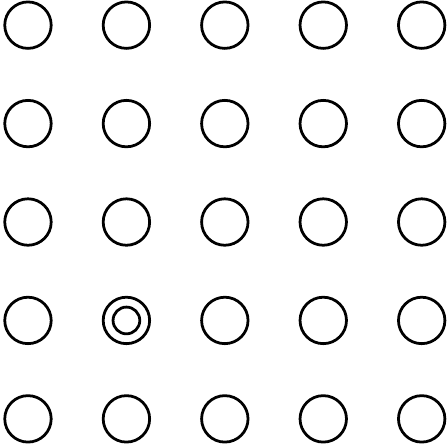} &
   \hskip 5mm\includegraphics[width=0.4\columnwidth]{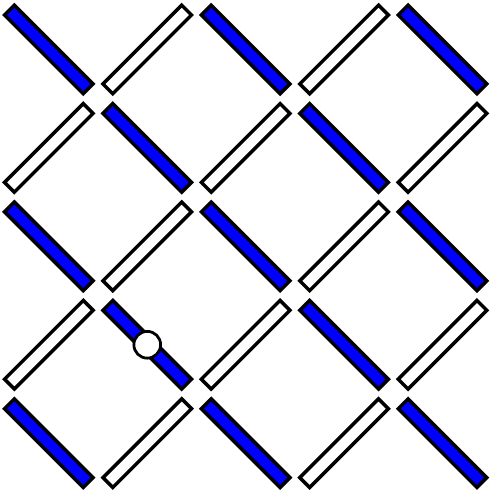} \\[1mm]
  (c) & \hskip 5mm (d)\\[2mm]
   \includegraphics[width=0.4\columnwidth]{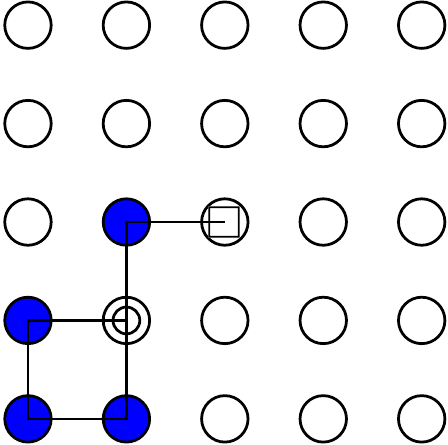} &
   \hskip 5mm\includegraphics[width=0.4\columnwidth]{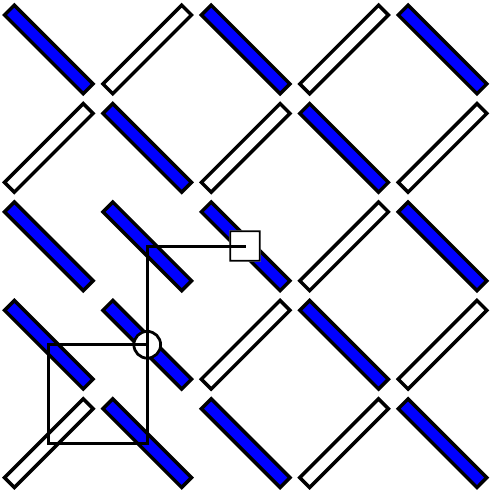} \\
 \end{tabular}
 \caption{\label{fig:FROE-FMCE}(Color online). (a) Initially ordered flipping
    rotator environment, $OFRE$. (b) Initially 
   checkerboard flipping mirror environment, $CFME$. (c) A walk on the $OFRE$
   at $t=6$. (d) A walk on the $CFME$ at $t=6$. The
   convention on the initial and final positions and initial velocity
   of both walkers and the scatterers is the same as in the previous
   Fig.}
  \end{center}    
\end{figure}

The walker moves with one of four velocities $\v{v}_0=(1,0)$,
$\v{v}_1=(0,1)$, $\v{v}_2=(-1,0)$, or $\v{v}_3=(0,-1)$ in discrete
time steps from one site on the environment to one of its nearest
neighbor sites according to his/her velocity and the state of the
scatterer he/she encounters.  The environment is defined by
$E=\{\sigma(x,y)\,|\,(x,y)\in\mathbb{Z}^2\}$ where
$\sigma(x,y)\,\in\,\{-1,1\}$ is the state of the scatterer at $(x,y)$.
When $\sigma(x,y)=1$ we say we have a right scatterer at $(x,y)$ and
when $\sigma(x,y)=-1$ a left scatterer. Environments of mirrors and
rotators will be denoted by $E_M$ and $E_R$ respectively in what
follows.  Right and left mirrors are shown schematically in
Figs.~\ref{fig:FMOE-FRCE} (a) and (c) and Figs. \ref{fig:FROE-FMCE}
(b) and (d). The reflection on the mirror forces the walker to turn an
angle of $\pm \pi/2$. A right (left) rotator scatters the walker an angle
of $\pi/2$ to his/her right (left) as we show in
 Figs.~\ref{fig:FMOE-FRCE} (b) and (d) and Figs. \ref{fig:FROE-FMCE}
(a) and (c).  After being scattered in either
environment at $(x,y)$, the walker moves to one of two neighboring
sites and $\sigma(x,y)$ flips by changing sign.

A walker with velocity $\v{v}_{k}$ is scattered with velocity
$\v{v}_{k'}$ with $k, k'=0,\dots,3$.  There are $4!$ different
injective scattering rules since if $\v{v}_k\neq \v{v}_l$ then
$\v{v}_{k'}\neq \v{v}_{l'}$. Of these, we chose those that scatter the walker
an angle of $\pm \pi/2$, that limits the scattering rules to four as we show in
Table~\ref{tab:square}.  The four rules of
Table~\ref{tab:square} are equivalent in couples so effectively,
there are only two injective scattering rules as we show next.  The
second and third columns of the Table show the scattering rule for
mirrors $M$. A walker moving horizontally to the right, $k=0$, will be
scattered vertically upwards when $\sigma=1$, $k'=1$, and vertically
downwards when $\sigma=-1$, $k'=3$.  A walker moving horizontally to
the left, $k=2$, will be scattered vertically downwards, when
$\sigma=1$ , $k'=3$, and vertically upwards when $\sigma=-1$,
$k'=1$. Thus, the first row of the scattering rule for a mirror fixes
the values of the third row. This is also valid for the second and
fourth rows of the scattering rule for $M$.

The fourth and fifth columns of Table~\ref{tab:square} show the
scattering rule for rotators $R$. A right rotator, $\sigma=1$,
scatters the walker to his/her right and a left rotator, $\sigma=-1$,
to his/her left. Again, the first and second rows fix the third
and fourth ones respectively.  The other two scattering rules, called
$A$ and $B$ are shown in the remaining columns of
Table~\ref{tab:square} where $A_{-1}$ and $A_1$ are the two states of
the $A$ scatterer and $B_{-1}$ and $B_1$ those of the $B$
scatterer. With $A_{-1}=-1$ and $A_1=1$, the $A$ scattering rule is the
same as that of the mirror $M$, and with $B_{-1}=-1$, $B_1=1$, the $B$
scattering rule is that of the rotator $R$. Hence, there are only two
injective rules that scatter the walker
by angles of $\pm\pi/2$.
\begin{table}
  \caption{\label{tab:square} Scattering rules for mirrors, $M$,
    and rotators, $R$, in $\mathbb{Z}^2$, $\v{v}_k\to\v{v}_{k'}$ with
    $k, k'=0,1,2,3$. The other two possible scattering rules are $A$
    and $B$, with $A_{-1}$, $A_1$, and $B_{-1}$, $B_1$ the orientations of
    the scatterers.}
 \vskip 2mm
   \begin{tabular}{|c|cc|cc|cc|cc|}
    \hline
    $k$ &\multicolumn{8}{c|}{$k'$}\\
    \hline
    &\multicolumn{2}{c|}{$M$}&\multicolumn{2}{c|}{$R$} 
    &\multicolumn{2}{c|}{$A$}&\multicolumn{2}{c|}{$B$}\\
    \hline
    & $\sigma=1$ & $\sigma=-1$ & $\sigma=1$ & $\sigma=-1$ 
    & \ $A_{-1}$ \ & \ $A_1$ \ & \ $B_{-1}$ \ & \ $B_1$ \  \\
    \hline
    0 & 1 & 3 & 3 & 1 & 3 & 1 & 1 & 3\\
    1 & 0 & 2 & 0 & 2 & 2 & 0 & 2 & 0\\
    2 & 3 & 1 & 1 & 3 & 1 & 3 & 3 & 1\\
    3 & 2 & 0 & 2 & 0 & 0 & 2 & 0 & 2\\
    \hline
 \end{tabular}
\end{table}

An initially ordered environment is one with
$\sigma(x,y)=1\,\,\forall\, (x,y)\in\mathbb{Z}^2$, or equivalently
$\sigma(x,y)=-1\,\,\forall\, (x,y)\in\mathbb{Z}^2$. In what follows
and without loss of generality we chose the first option. A
checkerboard environment is one in which at any site $(x,y)$, the
sign of the scatterers of the four nearest neighbor sites is
opposite to that at $(x,y)$.  Given the initial position of the walker
at the origin, there are two checkerboard environments, one with
$\sigma(0,0)=1$, the other one with $\sigma(0,0)=-1$.

The first example of this Sec. is that a walk initially at $(0,0)$
with velocity $(1,0)$ on an initially ordered flipping mirror
environment, $OFME$, is equivalent to a walk that starts with the same
initial position and velocity on an initially checkerboard flipping
rotator
environment, $CFRE$, provided this environment is chosen in such a way
that at time $t=0$ both walkers are scattered in the same direction.  The
second example is the equivalence of trajectories on an initially
ordered flipping rotator environment, $OFRE$, and on an initially
checkerboard flipping mirror environment, $CFME$, when the initial
position and velocity of both walkers satisfy the imposed conditions
of the first example.

In Figs.~\ref{fig:FMOE-FRCE} (a) and (b) we show a $OFME$ and a $CFRE$
respectively. The initial position of the walkers is at $(0,0)$ marked
by a small open circle. In Fig.~\ref{fig:FMOE-FRCE} (c) we show the
walk and the state of the mirror environment at time $t=6$, and in
Fig.~\ref{fig:FMOE-FRCE} (d) the walk and the state of the the rotator
environment at the same time.  Both walkers at $t=0$ have
$\v{v}_0=(1,0)$ and are scattered upwards as shown by the arrow from
$(0,0)$ to $(0,1)$. At $t=1$, both walkers are at $(0,1)$ and are
scattered to their right and at $t=2$ arrive at $(1,1)$.  Both walkers
are then scattered upwards and then horizontally to the right reaching
$(3,3)$ at $t=6$, shown by an open square.  Although the scatterers
flip after the walker passes, they do not influence the walk. Thus,
both walkers follow the same trajectory alternating between a vertical
and a horizontal step and moving diagonally with a speed of
$\sqrt{2}/2$.

In Figs.~\ref{fig:FROE-FMCE} (a) and (b) we show a $OFRE$ and a $CFME$
respectively. The initial position of the walkers on both environments
is shown by a small open circle at $(0,0)$.  In
Fig.~\ref{fig:FROE-FMCE} (c) we show the walk and the state of the
rotator environment at $t=6$ and in Fig.~\ref{fig:FROE-FMCE} (d) the
walk and the state of the mirror environment at the same time.  Both
walkers are at $(0,0)$ with velocity $\v{v}_0=(1,0)$ at $t=0$. In the
rotator environment, Fig.~\ref{fig:FROE-FMCE} (c), the walker is first
scattered to his/her right and moves downward reaching $(0,-1)$ at
$t=1$, is again scattered to his/her right to $(-1,-1)$ at $t=2$,
again to his/her right to $(-1,0)$ at $t=3$, and at $t=4$ is back at
the origin with velocity $\v{v}_0$. The same happens in the mirror
environment, Fig.~\ref{fig:FROE-FMCE} (d), the walker is scattered to
his/her right from $t=0$ to $t=4$ when he/she is back at the origin.
Since both scatterers at $(0,0)$ have flipped, both walkers will be
scattered to their left and move upwards and then they will scatter to
their right so that at $t=6$ are at $(1,1)$. Both walkers will scatter
to their right at sites that are visited for the first time, to
their left at sites that have been visited once, again
to their right at sites that have been visited two times
and so on.  Thus, their trajectories are equivalent up to time $t=6$.
In Fig.~\ref{fig:FM-FR} (b) we show the walk on a $OFRE$ for a longer
time.

In Theorem~\ref{thm:equivalence} we show that given a walk on one
environment, there is an equivalent walk on the other one.  Two walks
are equivalent if their paths $\{(x(t),y(t))| t\in\mathbb{N}\}$ are
the same.

\begin{theorem}\label{thm:equivalence}
  Let $E_M=\{\sigma_M(x,y)|(x,y)\in\mathbb{Z}^2\}$ be a mirror
  environment, $E_R=\{\sigma_R(x,y)|(x,y)\in\mathbb{Z}^2\}$ a
  rotator environment, and $d(x,y)=|x|+|y|$ for any
  $(x,y)\in\mathbb{Z}^2$. Given a walk with initial position $(0,0)$ and
  velocity $\v{v}_0$ on one environment, there is an equivalent walk
  with the same initial position and velocity on the other
  environment if
  \begin{equation}
   \label{eq:eqZ}
   \sigma_R(x,y)=(-1)^{d(x,y)+1}\sigma_M(x,y).
  \end{equation}  
\end{theorem}

\noindent {\it Proof of Theorem~\ref{thm:equivalence}.}
A site $(x,y)$ is even (odd) if $d$ is even (odd).  The number of
steps of any path from the origin to an even (odd) site is even (odd). Since
the initial velocity $\v{v}_0$ is horizontal, the velocity of the
walker at odd times will be vertical and at even times
horizontal, that is $\v{v}(2t+1)$ is either $\v{v}_1$ or $\v{v}_3$ and
$\v{v}(2t)$ is either $\v{v}_0$ or $\v{v}_2$ for $t=0,1,\dots$.  This
implies that the walker arrives to even sites with a horizontal
velocity, $\v{v}_0$ or $\v{v}_2$, and to odd sites with a vertical
velocity, $\v{v}_1$ or $\v{v}_3$.

From Table~\ref{tab:square} we get that if the velocity is
$\v{v}_1$ or $\v{v}_3$ the mirror and rotator scatterers of the same
sign act in the same way on the walker, but if the velocity is
$\v{v}_0$ or $\v{v}_2$ the mirror and rotator scatterers of the same
sign act in opposite ways on the walker. Thus to change an environment
of rotators for one of mirrors, or the inverse, in such a way
that the walkers follow the same trajectory, we have to
change the sign of the scatterers at even sites. This implies the
result. \hfill $\square$

If one of the environments is initially ordered, the other one is a
checkerboard according to the theorem.  We finish with a remark on the
initial conditions of the walker.  If the starting
point is $(x_0,y_0)$ the function $d(x,y)$ has to be replaced by the
function $d_{(x_0,y_0)}(x,y)=|(x-x_0)|+|(y-y_0)|$. This change implies
that the exponent of $-1$ in Eq.~\eqref{eq:eqZ} is replaced by
$d_{(x_0,y_0)}(x,y)+1$. If the initial velocity is changed to
$\v{v}_2$, the rule for changing the environment is the same as in
Theorem~\ref{thm:equivalence}, but if it is $\v{v}_1$ or $\v{v}_3$ the
exponent of $-1$ has to be replaced by $d_{(x_0,y_0)}(x,y)$ as can be
deduced from the proof above.


\section{Walks in ${\mathbb T}^2$}
\label{sec:triangle}

We consider walks on flipping environments in $\mathbb{T}^2$ where the
walker is scattered by angles of $\pm 2\pi/3$ with respect to his/her
velocity and show that there are only two injective scattering rules.
We then prove that given the initial position and velocity of a walker
in an environment with one type of scatterers, there is an environment
with the other type of scatterers such that the walks on the two
environments are equivalent.

\begin{figure}
  \begin{tabular}{cc}
  (a) & (b) \\
  \includegraphics[width=0.5\columnwidth]{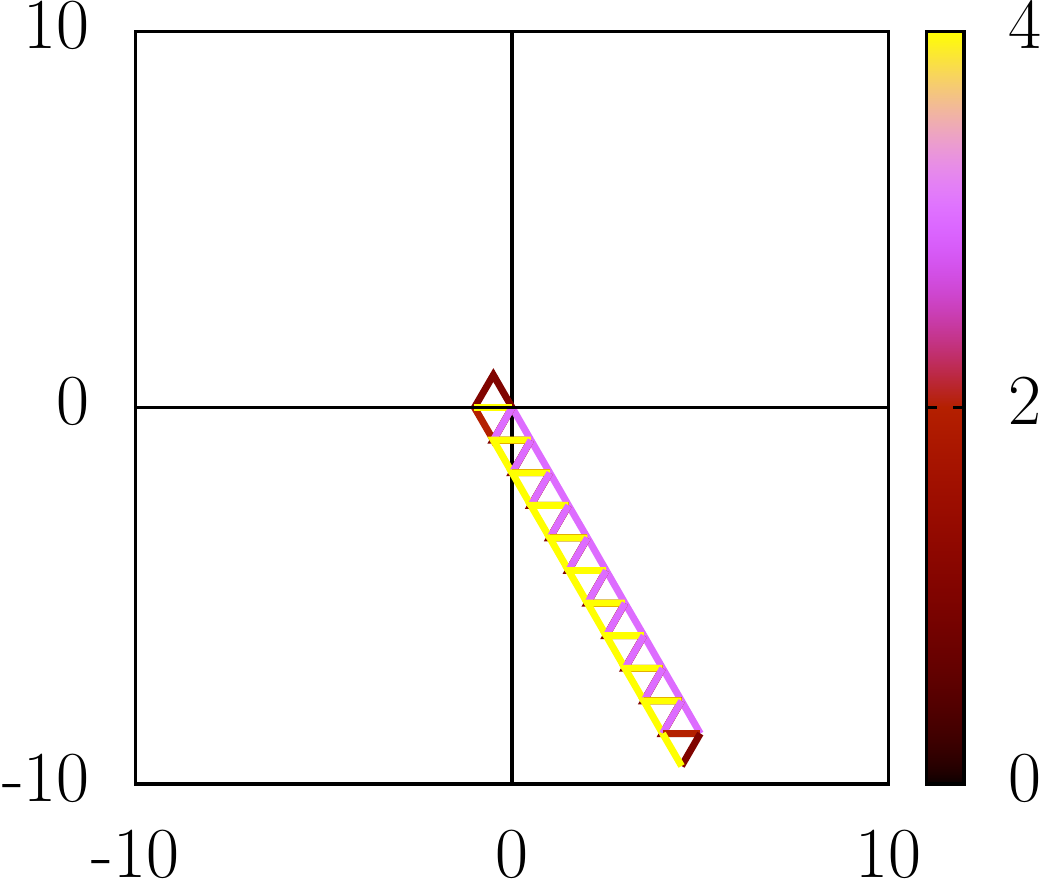} &
  \includegraphics[width=0.5\columnwidth]{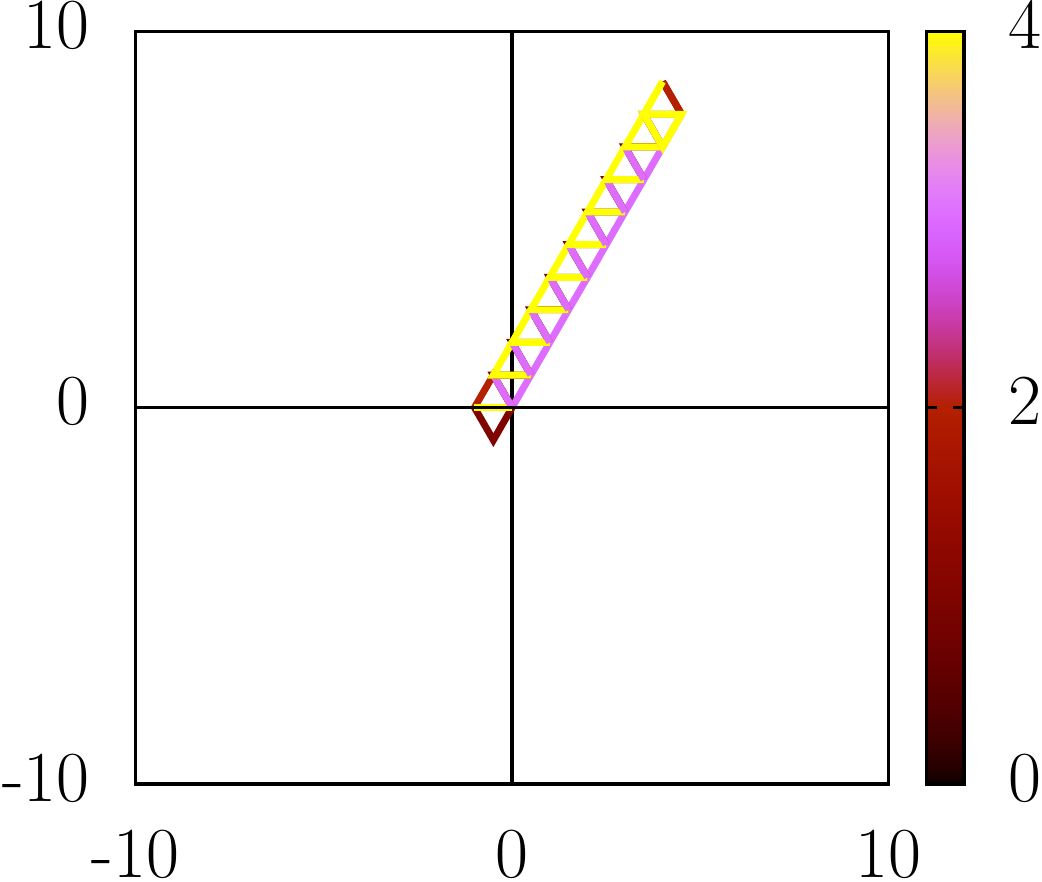}
  \end{tabular}
  \caption{\label{fig:triang-ordered} (Color online). Walks on
    initially ordered environments with with $\sigma(x,y)=1$ on all
    sites. (a) Flipping mirror environment $OFME$
    and (b) flipping rotator environment $OFRE$. In both cases, at $t=0$, the
    walker is at $(0,0)$ with velocity $\v{v}_0$. (The color scale
    indicates the number of times the walker is at a site.)}
\end{figure}

The six possible velocities in $\mathbb{T}^2$, with $h=\sqrt{3}/2$,
are $\v{v}_0=(1,0)$, $\v{v}_1=(1/2,h)$, $\v{v}_2=(-1/2,h)$,
$\v{v}_3=(-1,0)$, $\v{v}_4=(-1/2,-h,)$, and $\v{v}_5=(1/2,-h)$.  In
one time step, a walker with velocity $\v{v}_k$ is scattered
by angles
of
$\pm 2\pi/3$ with velocity $\v{v}_{k'}$ with respect to $\v{v}_k$ and
moves to one of two possible neighbor sites according to one of the
injective rules shown in Table~\ref{tab:triangular}. If $k$ is odd
(even), $k'$ is odd (even).  The scattering rules for mirrors and
rotators are shown in the columns $M$ and $R$ of
Table~\ref{tab:triangular}, respectively, corresponding to models 2B
and 1B of Ref.\cite{kong91a}. In the Table, the columns $A$ and $B$ are
the other two possible injective scattering rules. For the $A$ rule,
with $A_{-1}=-1$ and $A_1=1$ we obtain the mirror $M$ scattering rule,
and for the $B$ rule, $B_{-1}=-1$ and $B_1=1$ gives the rotator $R$
scattering rule. Thus there are only two injective scattering rules on
$\mathbb{T}^2$ for which the walker turns by angles of $\pm 2\pi/3$. In
Fig.~\ref{fig:triang-ordered} we show walks on initially ordered
flipping mirror and flipping rotator environments.  In both cases,
after a small transient, the walker advances half a site horizontally
and one site vertically every 8 time steps, moving with a speed of
$\sqrt{5}/16$ in the direction of $\v{v}_5$ in the mirror environment
and in the direction of $\v{v}_1$ in the rotator environment. On
initially disordered environments, the walker will also move in strips
as shown by Grosfils {\em et al}~\cite{grosfils99}.  The walks on
mirror and rotator environments are not
so strikingly different as those in $\mathbb{Z}^2$. In
Theorem~\ref{thm:triangularequivalence} we prove the equivalence of
walks on mirror and rotator environments.
\begin{table}
  \caption{\label{tab:triangular} Scattering rules for mirrors, $M$,
    and rotators, $R$, in $\mathbb{T}^2$, $\v{v}_k\to\v{v}_{k'}$ with
    $k, k'=0,\dots,5$. The other two possible scattering rules are $A$
    and $B$, with $A_{-1}$, $A_1$, and $B_{-1}$, $B_1$ the orientations of
    the scatterers.}
 \vskip 2mm
  \begin{tabular}{|c|cc|cc|cc|cc|}
    \hline
    $k$ &\multicolumn{8}{c|}{$k'$}\\
    \hline
    &\multicolumn{2}{c|}{$M$}&\multicolumn{2}{c|}{$R$}&\multicolumn{2}{c|}{$A$}&\multicolumn{2}{c|}{$B$}\\
    \hline
    & $\sigma=1$ & $\sigma=-1$ & $\sigma=1$ & \ $\sigma=-1$ \ & \ $A_{-1}$ \ &
    \ $A_1$ \ & \ $B_{-1}$ \ & \ $B_1$ \ \\
    \hline
    0 & 2 & 4 & 4 & 2 & 4 & 2 & 2 & 4 \\
    1 & 5 & 3 & 5 & 3 & 3 & 5 & 3 & 5 \\
    2 & 4 & 0 & 0 & 4 & 0 & 4 & 4 & 0 \\
    3 & 1 & 5 & 1 & 5 & 5 & 1 & 5 & 1 \\
    4 & 0 & 2 & 2 & 0 & 2 & 0 & 0 & 2 \\
    5 & 3 & 1 & 3 & 1 & 1 & 3 & 1 & 3 \\
    \hline
  \end{tabular}  
\end{table}  

\begin{theorem}\label{thm:triangularequivalence}
 Let $E_M=\{\sigma_M(x,y)|(x,y)\in\mathbb{T}^2\}$ be a mirror
 environment and $E_R=\{\sigma_R(x,y)|(x,y)\in\mathbb{T}^2\}$ a
 rotator environment.
 \renewcommand{\labelenumi}{(\alph{enumi})}
 \begin{enumerate}
 \item For every walk on one environment with some initial position
   and initial velocity $\v{v}_0=(1,0)$, $\v{v}_2=(-1/2,h)$ or
   $\v{v}_4=(-1/2,-h,)$ there is an equivalent walk on the other
   environment with the same initial position and velocity of the
   walker, if $\sigma_R(x,y)=-\sigma_M(x,y)\,\forall\,
   (x,y)\in\mathbb{T}^2$.
 \item For every walk on one environment with some initial position
   and initial velocity $\v{v}_1=(1/2,h)$, $\v{v}_3=(-1,0)$, or
   $\v{v}_5=(1/2,-h)$ there is an equivalent walk on the other
   environment with the same initial position and velocity of the
   walker, if $\sigma_R(x,y)=\sigma_M(x,y)\,\forall\,
   (x,y)\in\mathbb{T}^2$.
  \end{enumerate}
\end{theorem}
{\it Proof of Theorem~\ref{thm:triangularequivalence}.}
 \renewcommand{\labelenumi}{(\alph{enumi})}
 \begin{enumerate}
 \item Given that the walker has an initial velocity
   $\v{v}_k$ with $k$ even, the velocity at any time step will also
   have $k$ even on both environments as shown in
   Table~\ref{tab:triangular}. The columns in the rows of
   Table~\ref{tab:triangular} corresponding to $k$ even and the mirror
   scattering rule are inverted with respect to those of the rotator
   rule. Then walks on the mirror and the rotator environments with
   the same initial position and velocity $\v{v}_k$ are
   equivalent if $\sigma_R(x,y)=-\sigma_M(x,y)$.
 \item Given that the walker has an initial velocity $\v{v}_k$ with
   $k$ odd, the velocity at any time step will also have $k$ odd on
   both environments as shown in Table~\ref{tab:triangular}. The
   mirror and scattering rules in Table~\ref{tab:triangular} coincide
   for $k$ odd. Then walks on the two environments with the same
   initial position and velocity $\v{v}_k$, for $k$ odd,
   are equivalent if
   $\sigma_R(x,y)=\sigma_M(x,y)$.
 \end{enumerate}
 

 \section{Walks in ${\mathbb H}^2$}
\label{sec:hexagon}

As mentioned in Sec.~\ref{intro}, the equivalence of walks in
$\mathbb{H}^2$ was proven in Ref,~\cite{webb2014}. For completeness, we
present a proof of the same result.
A walker has a velocity
$\v{v}_k$, $k=0,\dots,5$ as in the previous Sec. and is scattered at
every time step by angles of $\pm\pi/3$.  Depending on the value of the
initial velocity, the walker will visit a different hexagonal two
dimensional lattice $\mathbb{H}^2$.  We show, as in the previous
Secs., that there are only two injective scattering rules, one gives
way to a rotator environment and the other one to a mirror
environment.  We show that walks on mirror and rotator environments in
$\mathbb{H}^2$ are qualitatively different and prove that for a walk
with fixed initial position and velocity on one environment, there is
an equivalent walk with the same initial conditions on the other
environment.

In Table~\ref{tab:hexagonal} we show the four injective scattering
rules.  A walker with velocity $\v{v}_k$ is scattered with velocity
$\v{v}_{k'}$.  If $k$ is even (odd), $k'$ is odd (even) for the four
rules.  In the mirror rule, shown in the $M$ columns of the Table, if
$\v{v}_k$ is scattered to $\v{v}_{k'}$, then $\v{v}_{k'}$ is scattered
to $\v{v}_k$ for the same value of $\sigma$.  This also happens for
walks in mirror environments in $\mathbb{Z}^2$.  The rotator rule is
shown in the $R$ columns of the Table, a walker with velocity
$\v{v}_k$ is scattered with velocity $\v{v}_{k'}$ with
$k'=(k-\sigma)\!\!\!\mod 6$. As in the previous cases the rules are
equivalent by couples. If $A_{-1}=-1$, $A_1=1$, we obtain the mirror
$M$ scattering rule, and if $B_{-1}=-1$, $B_1=1$, we obtain the
rotator $R$ scattering rule.
\begin{table}
  \caption{\label{tab:hexagonal} Scattering rules for mirrors, $M$,
    and rotators, $R$, in $\mathbb{H}^2$, $\v{v}_k\to\v{v}_{k'}$ with
    $k, k'=0,\dots,5$. The other two possible scattering rules are $A$
    and $B$, with $A_{-1}$, $A_1$, and $B_{-1}$, $B_1$ the orientations of
    the scatterers.}
 \vskip 2mm
  \begin{tabular}{|c|c c|c c|c c|c c|}
    \hline
    $k$ &\multicolumn{8}{c|}{$k'$}\\
    \hline
    &\multicolumn{2}{c|}{$M$}&\multicolumn{2}{c|}{$R$}&\multicolumn{2}{c|}{$A$}&\multicolumn{2}{c|}{$B$}\\
    \hline
    & $\sigma=1$ & $\sigma=-1$ & $\sigma=1$ & $\sigma=-1$ & \ $A_{-1}$ \ &
    \ $A_1$ \ & \ $B_{-1}$ \ & \ $B_1$ \ \\
    \hline
    0 & 5 & 1 & 5 & 1 & 1 & 5 & 1 & 5 \\
    1 & 2 & 0 & 0 & 2 & 0 & 2 & 2 & 0 \\
    2 & 1 & 3 & 1 & 3 & 3 & 1 & 3 & 1 \\
    3 & 4 & 2 & 2 & 4 & 2 & 4 & 4 & 2 \\
    4 & 3 & 5 & 3 & 5 & 5 & 3 & 5 & 3 \\
    5 & 0 & 4 & 4 & 0 & 4 & 0 & 0 & 4 \\
    \hline
  \end{tabular}  
\end{table}  

\begin{figure}
  \begin{tabular}{cc}
    (a) & (b) \\
    \includegraphics[width=0.45\columnwidth]{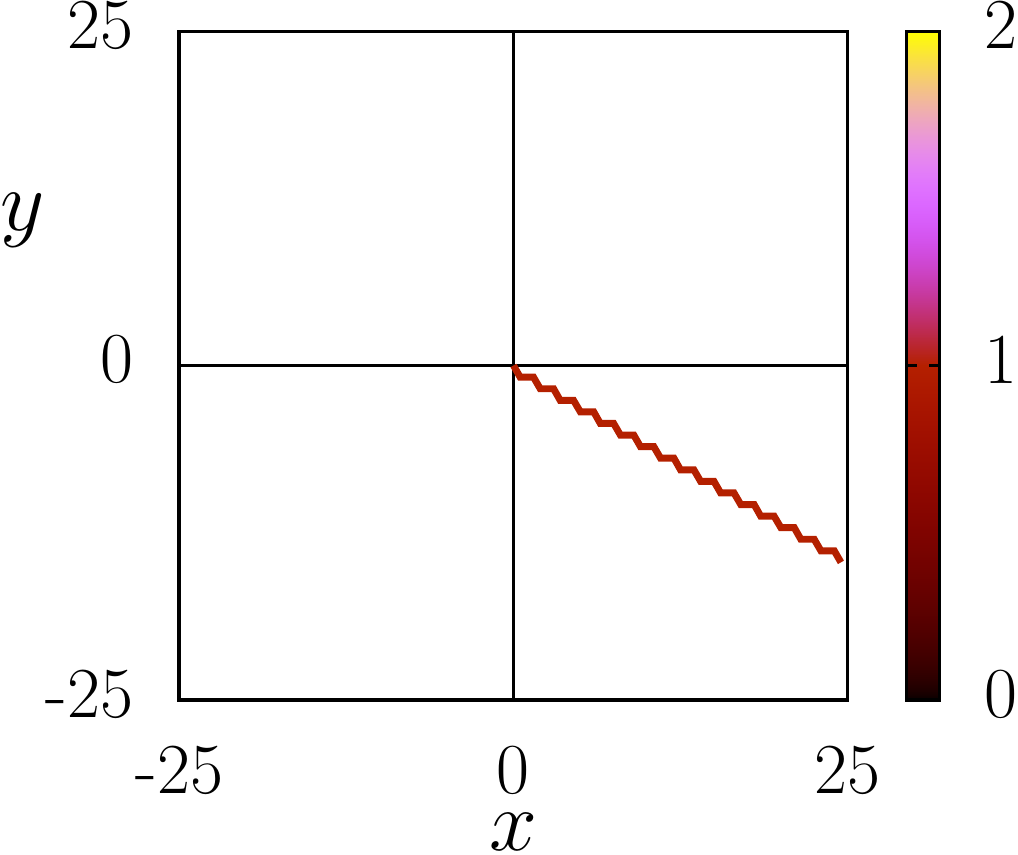} &
    \includegraphics[width=0.45\columnwidth]{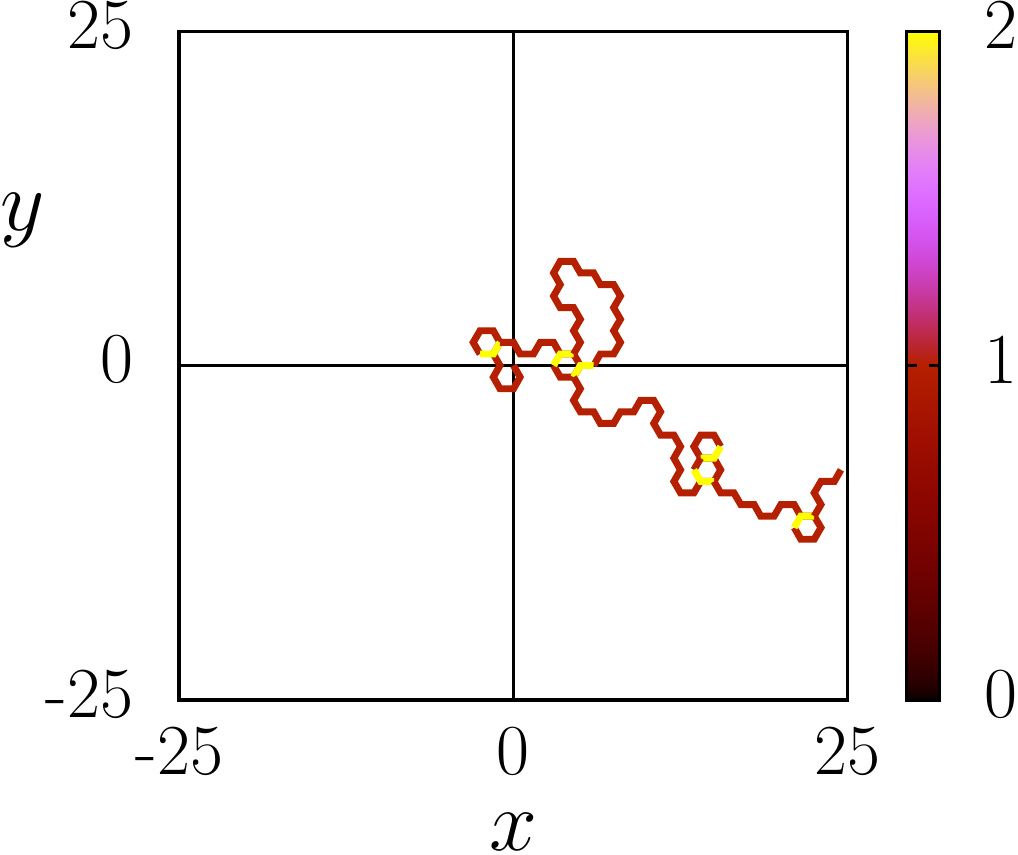} \\
    (c) & (d) \\
    \includegraphics[width=0.45\columnwidth]{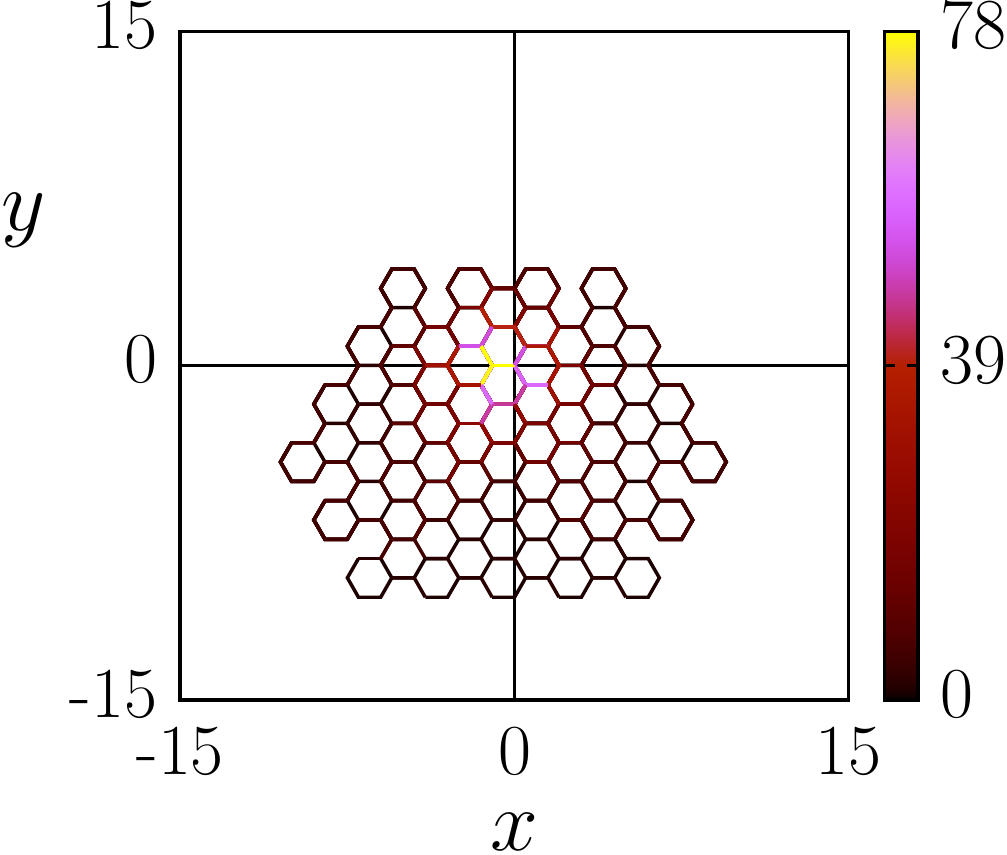} &
    \includegraphics[width=0.45\columnwidth]{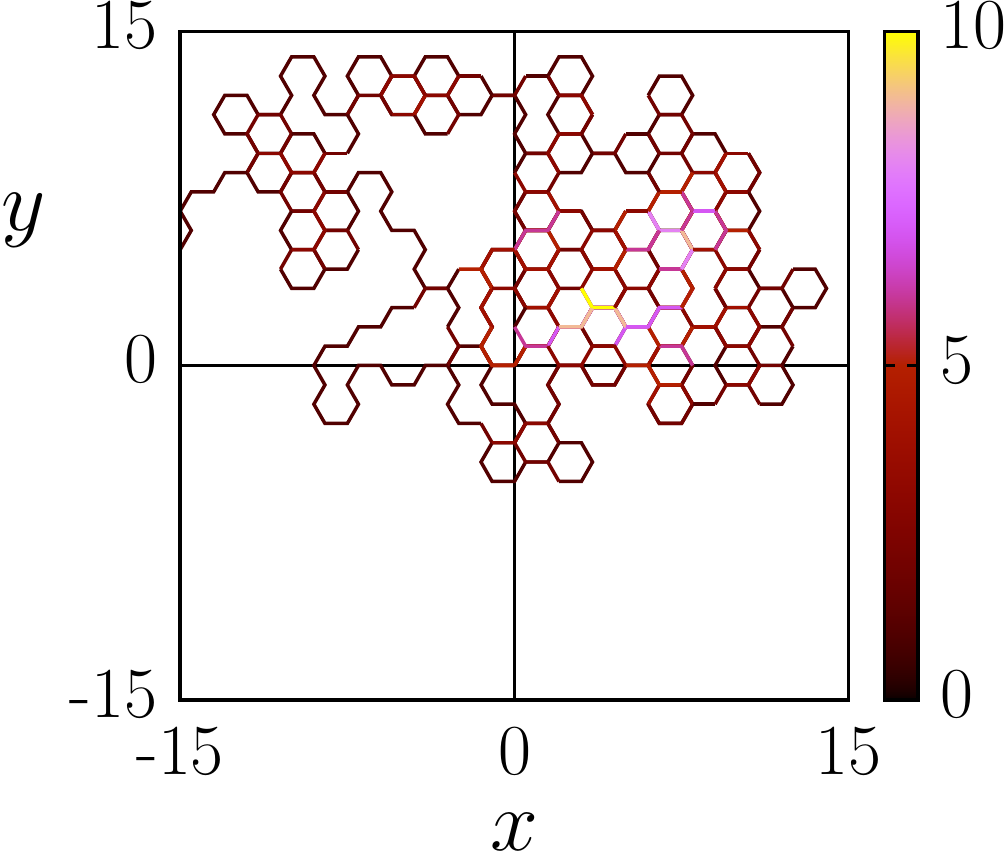} \\
 \end{tabular}
 \caption{\label{fig:mirror-rot-hex} (Color online.) (a) and (b)
   Trajectories of a walker on flipping mirror environments in
   $\mathbb{H}^2$. In (a), the environment is initially ordered with
   right mirrors. In (b), at $t=0$, the environment is
   ordered outside the square of side 40 and inside this
   square the probability that any given site has a right mirror is
   1/2. (c) and (d) Trajectories of a walker on flipping rotator environments
   in $\mathbb{H}^2$. In (c), the environment is initially ordered
   with right rotators. In (d), at $t=0$, the environment is
   ordered with right rotators outside the square of side
   40, and inside this square the probability that any given site has
   a right rotator is 1/2. In the four cases, the walker is initially
   at $(0,0)$ with velocity $\v{v}_0=(1,0)$. (The color scale
   indicates the number of times the walker is at a site.)  }
\end{figure}

In Fig.~\ref{fig:mirror-rot-hex} (a) we show a walk on an initially
ordered flipping mirror environment $OFME$ in $\mathbb{H}^2$. The walker
moves alternatively in directions $\v{v}_5$ and $\v{v}_0$ with a speed
of $\sqrt{3}/2$. When there is some disorder in the initial
environment of mirrors, the walker moves as we show 
in Fig.~\ref{fig:mirror-rot-hex} (b). A walk on an initially
ordered flipping rotator environment $OFRE$ is completely different
as we show in
Fig.~\ref{fig:mirror-rot-hex} (c), the walk is self-avoiding between returns
to the origin~\cite{webb2014}. In Fig.~\ref{fig:mirror-rot-hex} (d) we show
a walk on an initially disordered flipping rotator environment.

For a walk on one of the environments in $\mathbb{H}^2$, we
prove in Theorem~\ref{thm:hexagonalequivalence}
that there is an equivalent walk on the other environment.

\begin{theorem}\label{thm:hexagonalequivalence}
  Let $E_M=\{\sigma_M(x,y)|(x,y)\in\mathbb{H}^2\}$ be a mirror
  environment, $E_R=\{\sigma_M(x,y)|(x,y)\in\mathbb{H}^2\}$ a rotator
  environment, and $d(x,y)$ the number of steps of the shortest path
  (which is not in general unique) from $(0,0)$ to $(x,y)$.  
\renewcommand{\labelenumi}{(\alph{enumi})}
\begin{enumerate}
\item If
  \[
  \sigma_M(x,y)=(-1)^{d(x,y)}\sigma_R(x,y),
  \]
  walks that begin at $(0,0)$ with velocity $\v{v}_0=(1,0)$,
  $\v{v}_2=(-1/2,h)$ or $\v{v}_4=(-1/2,-h)$ on both
  environments are equivalent.  That is,
  $\{(x(t),y(t))| t\in\mathbb{N}\}$ is the same for both
  walks.
\item If
  \[
  \sigma_M(x,y)=(-1)^{d(x,y)+1}\sigma_R(x,y),
  \]
  walks that begin at $(0,0)$ with velocity $\v{v}_1=(1/2,h)$,
  $\v{v}_3=(-1,0)$ or $\v{v}_5=(1/2,-h)$ on both
  environments are equivalent. That is,
  $\{(x(t),y(t))| t\in\mathbb{N}\}$ is the same for both
  walks.
\end{enumerate}
\end{theorem}

\noindent {\it Proof of Theorem~\ref{thm:hexagonalequivalence}.}  If
$d$ is an even (odd) number, $(x,y)$ is an even (odd) site of
$\mathbb{H}^2$.  The nearest neighbors of an even (odd) site are odd
(even) sites. Note that if $(x,y)$ is an even (odd) site, any walk from the
origin to $(x,y)$ will visit an even (odd) number of sites.
From Table~\ref{tab:hexagonal}, a walk with initial velocity $\v{v}_k$
and $k$ even will have a velocity with $k$ odd at odd times and a
velocity with $k$ even at even times. Also from
Table~\ref{tab:hexagonal} we have that $M$ and $R$ scatterers with
the same sign of $\sigma$ scatter the walker in the same direction
$k'$ if $k$ is even and opposite directions if $k$ is odd.

Combining these two observations we obtain that if the initial
velocity has $k$ even, $\v{v}_0$, $\v{v}_2$ or $\v{v}_4$, the
scatterers at odd sites $(x,y)$ must have opposite signs in the two
environments in order to scatter the walker in the same direction.
This proves part (a) of the Theorem. Analogously, if the initial
velocity has odd $k$, $\v{v}_1$, $\v{v}_3$ or $\v{v}_5$, scatterers at
even sites have to have opposite signs in the $M$ and $R$
environments, proving part (b) of the Theorem. \hfill $\square$


\section{Concluding remarks}
\label{sec:conclusions}
 
We showed that on the three regular lattices on the plane, square,
triangular, and hexagonal, there are only two injective two state
scattering rules, rotators and mirrors.
We extended Webb and Cohen's
result of the equivalence of walks on flipping mirror and rotator environments in
hexagonal lattices to the equivalence of walks on triangular and
square lattices. The proofs of the equivalence of walks on both
environments are based on an interplay between the scattering
rules and the geometry of the lattice.

Given a walk on an initially ordered flipping rotator environment
$OFRE$ in $\mathbb{H}_2$, Cohen and Webb~\cite{webb2014} proved that
the walk is self-avoiding between successive returns to the origin. As
a consequence of the equivalence of walks, a walk on an initially
checkerboard flipping mirror environment $CFME$ in $\mathbb{H}_2$ will
be self-avoiding between successive returns to the origin.
In~\cite{webb2015}, Webb and Cohen study the trajectories on flipping
rotator environments. Starting with an initially ordered flipping
rotator environment (of right or left rotators), they prove that for a
walk in an initial environment obtained by changing the rotator at each
site with probability $p\in (0,1)$, then the trajectory starting at
$(0,0)$ with velocity $\v{v}_0$ will be periodic with probability
1. Applying Theorem~\ref{thm:hexagonalequivalence} we can state this
result for flipping mirror environments: starting with an initially
checkerboard flipping mirror environment, consider a walk in an initial
environment obtained by changing the mirror at each site with
probability $p\in (0,1)$, then the trajectory starting at $(0,0)$ with
velocity $\v{v}_0$ will be periodic with probability 1.

For any walk
on any environment with one type of scatterers in one of the three two
dimensional regular lattices, there is an equivalent walk on another
environment with the other type of scatterers. This means, that
whatever result is valid for walks with one type of scatterers is
valid for walks with the other type of scatterers if the second
environment is chosen according to the theorems proved above.


\section*{Acknowledgments}

It is a pleasure to acknowledge many enlightening discussions with
E. G. D. Cohen, H. Larralde, and M. L\'opez de Haro. Also, interesting
discussions with H.  D. Cort\'es Gonz\'alez and M. Valdez Gonz\'alez
are acknowledged. We also acknowledge the anonymous reviewers. Their
comments have enriched this presentation. A. Rechtman acknowledges the
support of LAISLA, the collaboration program between Mexico and France
for mathematics.

\bibliographystyle{unsrt} 
\bibliography{llg}

\begin{thebibliography}{10}

\bibitem{webb2014}
B.~{Z}. Webb and {E. G. D} Cohen.
\newblock Self-avoiding modes of motion in a deterministic {L}orentz lattice
  gas.
\newblock {\em J. Phys. A: Math. Theor}, 47:315202, 2014.

\bibitem{ehrenfest1912}
P.~Ehrenfest and T.~Ehrenfest.
\newblock Begriffliche {G}rundlagen der statistische {A}uffasung in der
  {M}echanik.
\newblock In {\em Encyklop\"adie der Mathematische Wissenschaften}, volume
  IV:2:II, No. 6. B. G. Teubner, (Leipzig), 1912.
\newblock Translated to English by M. J. Moravcsik in {\em The Conceptual
  Foundations of the Statistical Approach in Mechanics}, Cornell University
  Press, Ithaca NY, (1959).

\bibitem{bunimovich04}
L.~A. Bunimovich.
\newblock Deterministic walks in random environments.
\newblock {\em Physica D}, 187:20, 2004.

\bibitem{ruijgrok88}
Th.~W. Ruijgrok and E.~G.~D. Cohen.
\newblock Deterministic lattice gas models.
\newblock {\em Phys. Lett. A}, 133:415, 1988.

\bibitem{cohen92}
E.~G.~D. Cohen.
\newblock New types of diffusion in lattice gas cellular automata.
\newblock In M.Mareschal and B.~Holian, editors, {\em Microscopic Simulations
  of Complex Hydrodynamic Phenomena}, volume NATO ASI Series B, vol 292, 1992.

\bibitem{gunn85}
J.~M.~F. Gunn and M.~Ortuno.
\newblock Percolation and motion in a simple random environment.
\newblock {\em J. Phys. A}, 18:L1095, 1985.

\bibitem{langton86}
C.~G. Langton.
\newblock Studying artificial life with cellular automata.
\newblock {\em Physica D}, 22:120, 1986.

\bibitem{meng94}
Hsin-Fei Meng and E.~G.~D. Cohen.
\newblock Growth, self-randomization, and propagation in a {L}orentz lattice
  gas.
\newblock {\em Phys. Rev. E}, 50:2482, 1994.

\bibitem{kong91}
X.~P. Kong and E.~G.~D. Cohen.
\newblock Lorentz lattice gases, abnormal diffusion, and polymer statistics.
\newblock {\em J. Stat. Phys.}, 62:1153, 1991.

\bibitem{cohen95}
E.~G.~D. Cohen and F.Wang.
\newblock New results for diffusion in {L}orentz lattice gas cellular automata.
\newblock {\em J. Stat. Phys.}, 81:445, 1995.

\bibitem{cohen95a}
E.~G.~D. Cohen and F.Wang.
\newblock Novel phenomena in {L}orentz lattice gases.
\newblock {\em Physica A}, 219:56, 1995.

\bibitem{bunimovich92}
L.~A. Bunimovich and S.~E. Troubetzkoy.
\newblock Recurrence properties of {L}orentz lattice gas cellular automata.
\newblock {\em J. Stat. Phys.}, 67:289, 1992.

\bibitem{wang95}
F.~Wang and {E. G. D.} Cohen.
\newblock Diffusion in {L}orentz lattice gas cellular automata: the honeycomb
  and quasi-lattices compared with the square and triangular lattices.
\newblock {\em J. Stat. Phys}, 81:467, 1995.

\bibitem{webb2015}
B.~{Z}. Webb and {E. G. D} Cohen.
\newblock Self-limiting trajectories of a particle moving deterministically in
  a random medium.
\newblock {\em J. Phys. A: Math. Theor.}, 48:485203, 2014.

\bibitem{gale95}
D.~Gale, J.~Propp, and S.~Troubetzkoy.
\newblock Further travels with my ant.
\newblock {\em Math. Intelligencer}, 17:48, 1995.

\bibitem{boon01}
J.~P. Boon.
\newblock How fast does {L}angton's ant move?
\newblock {\em J. Stat. Phys.}, 102:355, 2001.

\bibitem{kong91a}
X.~P. Kong and E.~G.~D. Cohen.
\newblock Diffusion and propagation in triangular {L}orentz lattice gas
  cellular automata.
\newblock {\em J. Stat. Phys.}, 62:737, 1991.

\bibitem{grosfils99}
P.~Grosfils, J.~P. Boon, E.~G.~D. Cohen, and L.~A. Bunimovich.
\newblock Propagation and organization in lattice random media.
\newblock {\em J. Stat. Phys.}, 97:575, 1999.

\end{thebibliography}
\end{document}